\documentclass[twocolumn,showpacs,preprintnumbers,pra,showkeys]{revtex4}
\usepackage{graphicx}

\newcommand{\beq}{\begin{equation}}
\newcommand{\eeq}{\end{equation}}
\newcommand{\beqa}{\begin{eqnarray}}
\newcommand{\eeqan}{\end{eqnarray*}}
\newcommand{\beqan}{\begin{eqnarray*}}
\newcommand{\eeqa}{\end{eqnarray}}
\newcommand{\bra}[1]{\left\langle{#1}\right|}
\newcommand{\ket}[1]{\left|{#1}\right\rangle}

\begin{document}

\title{Experimental demonstration of quantum source coding}
\date{\today}

\author{
Yasuyoshi Mitsumori$^{1, 2}$,
John A. Vaccaro$^{3}$,
Stephen M. Barnett$^{4}$,
Erika Andersson$^{4}$,
Atsushi Hasegawa$^{1, 2}$,
Masahiro Takeoka$^{1, 2}$,
and
Masahide Sasaki$^{1, 2}$}
\address{
$^1$
    Communications Research Laboratory,Koganei, 
    4-2-1 Nukuikita, Koganei, Tokyo 184-8795, Japan\\
$^2$
    CREST, Japan Science and Technology Corporation, 
    3-13-3 Shibuya, Tokyo 150-0002, Japan\\
$^3$
    Department of Physical Science,
    University of Hertfordshire,
    College Lane Hatfield AL109AB, U.K.\\
$^4$
    Department of Physics and Applied Physics,
    University of Strathclyde, Glasgow G4 0NG, Scotland}
\email{J.A.Vaccaro@herts.ac.uk}
\email{steve@phys.strath.ac.uk}
\email{psasaki@crl.go.jp}

\begin{abstract}
We report an experimental demonstration of Schumacher's quantum
noiseless coding theorem. Our experiment employs a sequence of
single photons each of which represents three qubits. We initially
prepare each photon in one of a set of 8 non-orthogonal codeword
states corresponding to the value of a block of three binary
letters. We use quantum coding to compress this quantum data into
a two-qubit quantum channel and then uncompress the two-qubit
channel to restore the original data with a fidelity approaching
the theoretical limit.
\end{abstract}


\maketitle


\noindent {\bf Introduction}\\
 \noindent
The coding of messages is a fundamental issue in information
theory. There are two basic coding problems, namely \textit{how to
represent messages as efficiently as possible} and \textit{how to
transmit messages as precisely as possible}. The former is called
\textit{source coding}, and is related practically to data
compression, while the latter is called \textit{channel coding}
and is concerned with error correction. All information processing
techniques are connected with these two kinds of coding problem.
We focus on source coding in this report.

Essentially, source coding entails the coding of common alphabet in
a message as short sequences of code letters, such as the binary
digits $\{0,1\}$, and uncommon alphabet as longer sequences, to
make the average length of the coded message as short as possible.
The unequal frequencies of the letters imply a redundancy that
enables the compression of the message. Shannon's source coding
theorem gives the bounds on the degree a classical message can be
compressed. For a source of alphabet $\{A, B, \ldots, Z\}$ with
given prior probabilities $\{ P(A), P(B), \ldots, P(Z) \}$, the
minimum average length of the coded message is given by the
Shannon entropy
\begin{equation}
H = -\sum_{\mskip-100mu n=A,B,\ldots\mskip-100mu}
   P(n) \log_2 P(n)
\,.
\label{H(X)}
\end{equation}
$H$ takes its maximum value when all alphabet appear with equal
probability, that is, when we know nothing better than a random
guess for each element. Then any compression is impossible.

In quantum domain, there is another kind of redundancy when the
letters are conveyed by the non-orthogonal quantum states,
$\ket{\psi_A}$, $\ket{\psi_B}$, $\ket{\psi_C}$, $\cdots$ with
corresponding probabilities  $P_A$, $P_B$, $P_C$, $\cdots$.
Significantly, compression is possible here even if
$P_A=P_B=P_C=\cdots$, in contrast to the classical case. Recently
Schumacher and Jozsa derived the quantum version of the source
coding theorem. The {\em quantum noiseless coding theorem}
\cite{Schumacher95,Jozsa_Schumacher94} implies that by coding the
quantum message in  blocks of $K$ letters, $KS(\hat\rho)$ qubits
are necessary to encode each block in the limit $K\to\infty$,
where $S(\hat\rho)$ is the von Neumann entropy of the density
operator $\hat\rho=\sum P_n\ket{\psi_n}\bra{\psi_n}$ representing
the average state of the letter states.

In addition to its central role in quantum information theory, the
compression of non-orthogonal data sets has significant practical
advantages.  For example, in long-haul optical communication
channels one must deal with sequences of attenuated weak coherent
pulses, that is, non-orthogonal states. Expensive quantum channel
resources can be saved by compressing the sequences before storing
or relaying to another channel.

Given its fundamental as well as practical importance, it is
perhaps surprising that quantum source coding has not been
demonstrated experimentally to date. We report an experimental
demonstration of the reliable communication of 3-qubit codewords
over a 2-qubit quantum channel. The minimum resources needed for
an analogous classical channel would be 3 bits per codeword.\\


\noindent
{\bf Quantum coding protocols}\\
 \noindent
Our demonstration is based on the example given by Jozsa and
Schumacher \cite{Jozsa_Schumacher94}.  Imagine Alice needs to send
Bob a message composed of an alphabet of 2 letters, ``$+$'' and
``$-$'', represented by the letter states $\ket{\psi_+}$ and
$\ket{\psi_-}$,
\beq
   \ket{\psi_\pm} = \alpha\ket{0} + \beta_\pm\ket{1}\ .
\eeq
Here $\ket{0}$, $\ket{1}$ are an orthonormal (computational)
basis, $\beta_{\pm}=\pm\beta$, $\alpha^2+\beta^2=1$, and for
clarity we assume $\alpha$ and $\beta$ are real numbers. Let the
letter states occur with equal likelihood so that the density
operator representing the average letter state is
$\hat{\rho}=\alpha^2\ket{0}\bra{0}+\beta^2\ket{1}\bra{1}$.  The
corresponding von Neumann entropy is
$S(\hat\rho)=-\alpha^2\log_2\alpha^2-\beta^2\log_2\beta^2$.

If the letter states are orthogonal,
$\alpha^2=\beta^2=\frac{1}{2}$, then 1 qubit (or classically 1
bit) is needed to encode each letter faithfully. In this case a
sequence of letter states cannot be compressed to a smaller code.
However, the von Neumann entropy of $\hat{\rho}$ is 0.4690 bits
for the case $\alpha^2=0.9$ \cite{Jozsa_Schumacher94}. According
to the quantum noiseless coding theorem, in the limit of large
block sizes Alice needs approximately $1/2$ qubit per letter state
to faithfully transmit the message to Bob.

Following \cite{Jozsa_Schumacher94} we use blocks of 3 letter
states:
%
%
\beqa
   \ket{B_{\rm\bf L}} &=&
   \ket{\psi_{L_1}}\otimes\ket{\psi_{L_2}}\otimes\ket{\psi_{L_3}}
   \nonumber\\
   &=&
   \alpha^3\ket{000}
   +\alpha^2\left(
             \beta_{L_1}\ket{100}
            +\beta_{L_2}\ket{010}
            +\beta_{L_3}\ket{001}
            \right)
   \nonumber\\
   &&+\alpha\left(
             \beta_{L_1}\beta_{L_2}\ket{110}
            +\beta_{L_2}\beta_{L_3}\ket{011}
            +\beta_{L_1}\beta_{L_3}\ket{101}
            \right)
   \nonumber\\
   &&+\beta_{L_1}\beta_{L_2}\beta_{L_3}\ket{111}
\eeqa
where ${\rm\bf L}=(L_1,L_2,L_3)$ and $L_1$, $L_2$ and
$L_3\in\{+,-\}$.  The index ${\rm\bf L}$ selects one of 8 possible
letter state configurations.
In our quantum coding scheme \cite{differ}, Alice first applies
the unitary transformation $\hat{U}$ which leaves all
computational bases states unchanged except for the following
mapping $\hat{U}\ket{100} = \ket{011}$ and $\hat{U}\ket{011}=
\ket{100}$. 
The state of a block after the application of $\hat{U}$ is
\beq
   \hat{U}\ket{B_{\rm\bf L}}
       =\alpha^2\sqrt{1+2\beta^2}\ket{0}\otimes\ket{\mu_{\rm\bf L}}
       +\beta^2\sqrt{1+2\alpha^2}\ket{1}\otimes\ket{\nu_{\rm\bf L}}
\eeq
where
\beqa
   \ket{\mu_{\rm\bf L}}
   &=&\frac{1}{\sqrt{1+2\beta^2}}
      \left(
       \alpha\ket{00}
      +\beta_{L_1}\ket{11}
      +\beta_{L_2}\ket{10}
      +\beta_{L_3}\ket{01}\right)\nonumber\\
   \\
   \ket{\nu_{\rm\bf L}}
   &=&\frac{1}{\beta^2 \sqrt{1+2\alpha^2}}
       \left[
         \alpha
            \left(
               \beta_{L_1}\beta_{L_2}\ket{10}
              +\beta_{L_1}\beta_{L_3}\ket{01}
                   \right.\right.\nonumber\\
               & &\ \left.\left.
              +\beta_{L_2}\beta_{L_3}\ket{00}
            \right)
      +\beta_{L_1}\beta_{L_2}\beta_{L_3}\ket{11}
      \right]\ .
\eeqa
Alice then makes a projection measurement of the first (leftmost)
qubit in the computational basis.  The last two qubits represent
the coded block state sent to Bob.
We consider two different protocols corresponding to two different
actions Alice takes when the projective measurement results in the
state $\ket{1}$.
Essentially, the coding protocols amount to a perfect transmission of the 
most likely parts of letter state configurations, and a less faithful 
transmission of, or even discarding of, the remaining less likely part.

The first protocol, which we shall label P$_1$, is to treat the
projection measurement result $\ket{1}$ as a failure. Under this
protocol the state of the 2-qubit quantum channel is
\beq
    \hat{\rho}_{\rm\bf L}^{(1)} = \ket{\mu_{\rm\bf L}}\bra{\mu_{\rm\bf L}}
\eeq
%
%
with probability $p=\alpha^4(1+2\beta^2)$ and a state of zero
overlap with any block state with probability $1-p$. Bob decodes
the state $\hat{\rho}_{\rm\bf L}^{(1)}$ at his end of the quantum
channel by preparing an extra qubit in the state $\ket{0}$ and
applying the inverse of $\hat{U}$; this results in the decoded
state $\hat{\Phi}_{\rm\bf L}^{(1)} = \hat{U}^\dagger \left(
\ket{0}\bra{0}\otimes\hat{\rho}_{\rm\bf L}^{(1)} \right) \hat{U}\
$. The fidelity of the whole quantum coding-decoding operation for
P$_1$ is given by
\beq
  F_1 = \sum_{\rm\bf L}\frac{1}{8}\bra{B_{\rm\bf L}}\hat{\Phi}_{\rm\bf L}^{(1)}\ket{B_{\rm\bf L}}
      =\alpha^8(1+2\beta^2)^2\ .
\eeq
$F_1$ is plotted as the solid curve in Fig.\ 1 and has a value of
$0.9448$ at $\alpha^2=0.9$.

The second protocol, P$_2$, yields a higher fidelity than that of
P$_1$. In this case Alice prepares the quantum channel in the
state $\ket{00}$ in the event that her projection measurement
results in the state $\ket{1}$. This operation results in the
average state of the quantum channel as
\beq
    \hat{\rho}_{\rm\bf L}^{(2)}
        = \alpha^4(1+2\beta^2)\ket{\mu_{\rm\bf L}}\bra{\mu_{\rm\bf L}}
         +\beta^4(1+2\alpha^2)\ket{00}\bra{00}\ .
\eeq
Bob again adds an extra qubit in the state $\ket{0}$ and applies
the inverse operation $\hat{U}^\dagger$ to produce state
$\hat{\Phi}_{\rm\bf L}^{(2)}
=\hat{U}^{\dagger} \left( \ket{0}\bra{0}\otimes\hat{\rho}_{\rm\bf
L}^{(2)} \right) \hat{U}\ $ which has a corresponding fidelity of
\beq
   F_2 = \alpha^8(1+2\beta^2)^2+\alpha^6\beta^4(1+2\alpha^2)\ .
\eeq
The value of $F_2$ is plotted as the dashed curve in Fig.\ 1.
$F_2$ has a value of $0.9652$ at $\alpha^2=0.9$.

Finally, Jozsa and Schumacher also considered the simple protocol,
P$_3$, where Alice discards the state of every third letter and
encodes the remaining letters in a block of 2 qubits, and Bob
generates the state $\ket{0}$ for the missing letter state.  This
protocol yields an average fidelity of
\beq
  F_{3}=\alpha^2\ ,
\eeq
which is plotted as the dotted curve in Fig.\ 1.\\

\begin{figure}
\includegraphics[width=70mm]{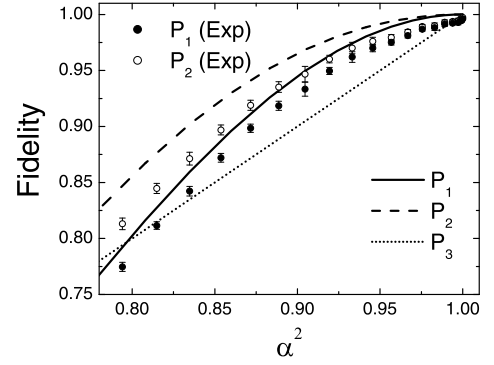}   %
\caption{Comparison of the fidelities of various protocols as a
function of the parameter $\alpha^2$.
\label{figLSdiagram}}
\end{figure}

\begin{figure}
\includegraphics[width=80mm]{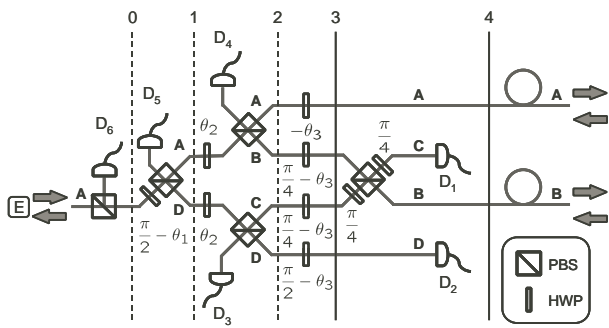}   %
\caption{Linear optics implementation of  three-qubit block coding
and decoding. \label{figOptCircuit}}
\end{figure}


\noindent
{\bf Optical scheme}\\
 \noindent
Fig.~\ref{figOptCircuit} shows  an ideal single-photon linear
optics implementation of the block coding scheme of the previous
section using polarizing beam splitters, $\lambda/2$ wave plates
and photodetectors.  One photon is used to represent the 3 qubits
in terms of two location qubits (the first 2 qubits) and a
polarization qubit \cite{LinearOptGating} (the third qubit).
The paths labeled A, B, C and D represent the states $\ket{00}$,
$\ket{01}$, $\ket{10}$ and $\ket{11}$ of the first two qubits,
respectively, and the polarization directions in the plane and
perpendicular to the plane of the optical circuit represent the
states $\ket{0}$ and $\ket{1}$ of the last qubit, respectively. We
parameterize the $n$th letter state by the angle
$\theta_n=\frac{1}{2}\arcsin(\beta_{L_n})$. The orientation of the
fast axis of each wave plate to the vertical direction is given
beside the wave plate in the figure.

First we discuss the preparation of the 3-qubit block state
$\ket{B_{\rm\bf L}}$. A horizontally-polarized photon enters the quantum
circuit at position E. The first location qubit is prepared by the
$\lambda/2$ wave plate and the polarization beam splitter between
vertical lines labeled 0 and 1 in terms of angle $\theta_1$.
Similarly, the second location qubit and the polarization qubit
are prepared between the vertical lines labeled 1 and 2, and 2 and
3, respectively. (Note that the photodetectors D$_3$, D$_4$, D$_5$
and D$_6$ are not used in state preparation.) The fully
state-prepared block appears as a photon in a superposition of 4
path and two linear polarization modes along the vertical line
labeled 3. This is the 3-qubit message Alice wants to compress and
communicate to Bob.

Next we discuss the quantum coding which takes place between
vertical lines 3 and 4. The unitary transformation $\hat{U}$ is
performed by the polarization beam splitter and the two
$\lambda/2$ wave plates in this section. The projection
measurement of the first qubit is performed by the photodetectors
D$_1$ and D$_2$ where the detection or non detection of a photon
projects the state onto $\ket{1}$ or $\ket{0}$, respectively. The
circuit in Fig.~\ref{figOptCircuit} implements protocol P$_1$
explicitly: if the photodetectors D$_1$ and D$_2$ detect a photon
no quanta will be present in the quantum channel and so the coding
results in failure.  The projective measurement is destructive in
this case. Protocol P$_2$ can be implemented by switching a
horizontally-polarized single photon source into the optical path
A to encode the state $\ket{000}$ each time one of the
photodetectors D$_1$ or D$_2$ detects a photon.
The encoded (compressed) 2-qubit message appears along vertical
line 4. This is transmitted to Bob.

The decoding of the quantum channel AB at Bob's site requires a
mirror image of the quantum circuit in Fig.\ \ref{figOptCircuit}
between lines 3 and 4 but without the photodetectors D$_1$ and
D$_2$. Moreover, the mirror image of the circuit to the left of
the line 3 can be used to determine the fidelity of the decoded
block message. The fidelity test results in a `yes'-`no' answer
for each coded-decoded block state as follows. The `yes' answer
(i.e. perfectly reconstructed letter block) is indicated by the
horizontally-polarized photon emerging from the mirror image of
point E. A `no' answer is indicated by the photon being detected
by one of the photodetectors D$_3$, D$_4$, D$_5$ or D$_6$ in the
mirror
image circuit \cite{7outcomes}. \\

\begin{figure}
\includegraphics[width=80mm]{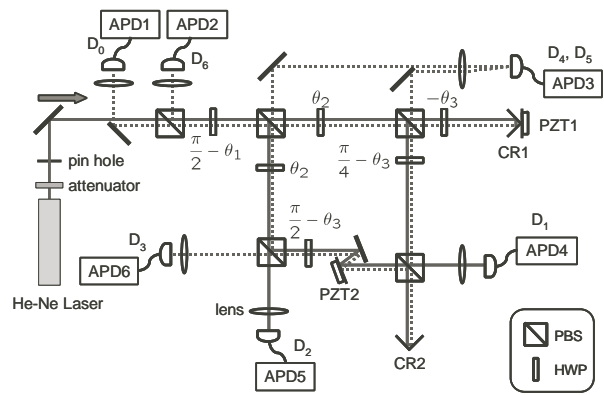}   %
\caption{Experimental setup.
\label{figExpSetup}}
\end{figure}


\noindent
{\bf Experimental Implementation}\\
 \noindent
Our actual experimental circuit is shown in
Fig.~\ref{figExpSetup}. Again, the orientation angle $\theta_{n}$
of each $\lambda/2$ wave plate is given in the figure. For
practical convenience, we did not construct an additional
mirror-image circuit for the decoding and fidelity check. Instead
we use corner reflectors (CR1 and CR2) to reflect the light in the
quantum channel back through the circuit (shown as dotted lines in
the figure) so that the coding and state-preparation circuits
operate as decoding and state-measurement circuits for the
reflected light. We use strongly attenuated light from a He-Ne
laser (wavelength 632.8 nm) as our single photon source. The CW
laser output of 1 mW power is attenuated to $\approx$50~fW which
corresponds to an average photon flux of 10$^5$ photons/sec.  The
average time between photons through our experiment far exceeds
the time taken for light to pass through the circuit
($\approx10^{-8}$\ s).

We use multimode optical fibers with coupling efficiency of more
than 80\% to direct the photons exiting the circuit to silicon
avalanche photodiodes (APDs). The quantum efficiency and dark
count of the APDs are typically 70\% and less than 100 counts/sec,
respectively. The labels for each APD (D$_1$--D$_6$) correspond to
those of photodetectors in Fig.~\ref{figOptCircuit} and the APD
labeled D$_0$ detects the `yes' answer of the fidelity test. Since
we do not need to discriminate the photodetection between D$_4$
and D$_5$, we use one APD for these detectors.

\begin{figure}
\includegraphics[width=90mm]{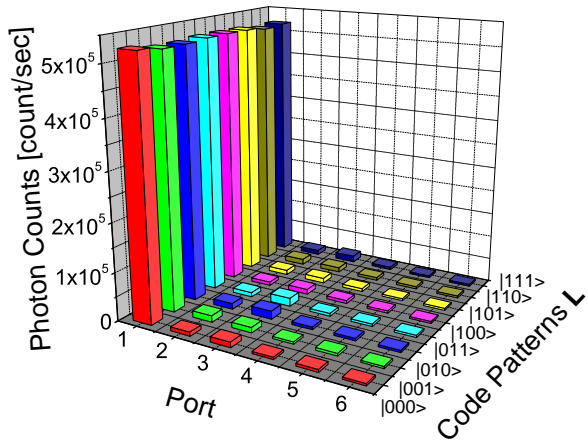}   %
\caption{Histogram of photon counts
for each block state ${\rm\bf L}$.
The number in the vertical axis labels six APDs from APD1 to APD6.
The parameter of the letter states and the fidelity are
$\alpha^2=0.9046$ and $F=0.933\pm0.006$, respectively.
The gating time of APDs is 5 sec.
\label{figHistogram}}
\end{figure}

In essence, the optical circuit consists of a Michaelson and a
Mach-Zehnder interferometers controlled by Piezo transducers PZT1
and PZT2, respectively. We use a bright reference light and adjust
the voltages of PZT1 and PZT2 to produce visibilities of more than
98\% for these interferometers. The reference light is then
switched off and the signal photons are guided into the circuit.
The single photon events are counted
by six APDs for each block state $|B_{\rm\bf L}\rangle$. The
gating time of the APDs is 5 sec with the combined count over 1
sec being of the order of 10$^5$. Since the whole apparatus is
shielded by a black box, the number of background photons is much
smaller than
the dark count of the APDs.  Also we estimate the number of events
where two photons are present simultaneously in the circuit to be
less than half the dark count.

Our use of a photon source with random arrival times means that
the quantum coding-decoding operations occur in the context of
post-selection measurements; that is, we know that a quantum
coding-decoding operation has taken place {\it after} it has
occurred, and, due to the limited efficiency of the
photodetectors, in a subset of possible cases.

The experimental fidelity for protocol P$_1$ is given by
\begin{equation}
  \label{eqExpF2}
   F_1^{\rm ex} = \sum_{\rm\bf L} \frac{1}{8}
       \frac{N_0^{\rm\bf L}}{\sum_{j=0}^{6} N_j^{\rm\bf L}}
\end{equation}
where $N_j^{\rm\bf L}$ is the number of photons detected by the
detector D$_j$ for the block state $|B_{\rm\bf L}\rangle$.
As an example,
Fig.~\ref{figHistogram} shows the photon counting data for the
letter state with $\alpha^2=0.9046$ for which the fidelity 
$F=0.933\pm0.006$.
%
By varying the angle $\theta$ the fidelity of our quantum
coding-decoding experiment can be compared by the theoretical
predictions given of the previous section over a range of $\alpha$
values. The results are shown in Fig.~\ref{figLSdiagram} as solid
circles.

For protocol P$_2$, rather than switching a horizontally-polarized
light source into channel A each time one of the photodetectors
D$_1$ or D$_2$ records a photon, we perform a 2 step procedure as
follows. The first step is the same as for protocol P$_1$ and, in
fact, we use the same photon counting data $N_j^{\rm\bf L}$ as
described. The second step corresponds to the transmission of a
horizontally-polarized photon in channel A for each of the photons
detected by D$_1$ and D$_2$ in the first step. For this purpose,
the corner reflector CR1 is removed and horizontally-polarized and
attenuated light from a He-Ne laser is directed into the circuit.
The number of photons used (i.e. the total number of photons
detected by all APDs) in this second step is adjusted to be
$N_1^{\rm\bf L}+N_2^{\rm\bf L}$ for each corresponding block state
$\ket{B_{\rm\bf L}}$. We can do this adjustment with an accuracy
of $\pm3$\% by carefully controlling the gating time of the APDs.
The total fidelity for this protocol is calculated as follows:
\begin{equation}
  \label{eqExpF3}
  F_2^{\rm ex} = \sum_{\rm\bf L} \frac{1}{8}
  \frac{N_0^{\rm\bf L} + N_0^{{\rm\bf L}\,(2)}}{\sum_{j=0}^{6} N_j^{\rm\bf L}}.
\end{equation}
where $N_0^{{\rm\bf L}\,(2)}$ is the total number of photons
detected by D$_0$ in the second step. We obtain the fidelities
corresponding to several $\alpha$ values and plot them as open
circles in Fig.~\ref{figLSdiagram}.  The experimental fidelities
for both protocols exceed that of the simple protocol.\\


\noindent
{\bf Discussion}\\
  \noindent
A message of equal-likely letters is not compressible classically.
In contrast, quantum source coding allows a quantum message of
equally likely (but non orthogonal) letter states to be compressed
\cite{Schumacher95,Jozsa_Schumacher94}.  Our compression of
3-qubit codewords gives a clear demonstration of this fundamental
principle.

The practical application of quantum source coding faces several
challenges. An immediate task is to demonstrate quantum source
coding using a single-photon-on-demand source. Another is to
replace spatial mode qubits with frequency mode qubits as this
leads to compression of bandwidth of quantum carrier. Of direct
practical importance is to demonstrate the coding and decoding of
a source of weak coherent states. Quantum circuits based on
measurement induced non-linearities and non-classical light
sources have potential in this regard. Our experiment is the first
step towards realizing these practical goals.

\acknowledgements
This work was supported by the British Council, the Royal Society of 
Edinburgh, the Scottish Executive Education and Lifelong Learning 
Department and the EU Marie Curie Fellowship program.



\begin{thebibliography}{99}
\bibitem{differ}
We perform the unitary transformation before the projection
operation, otherwise our scheme is identical to that of
\cite{Jozsa_Schumacher94}.
\bibitem{7outcomes}
We note that the photons arriving at D$_6$ have only vertical
polarization whereas those arriving at D$_3$ to D$_5$ can have
either vertical or horizontal polarization giving, in principle, 7
distinguishable outcomes for a `no' answer.
\end{thebibliography}
\end{document}